# Hyperincursive *Cogitata* and Incursive *Cogitantes*: Scholarly Discourse as a Strongly Anticipatory System

Loet Leydesdorff

University of Amsterdam, Amsterdam School of Communication Research (ASCoR), Kloveniersburgwal 48, 1012 CX Amsterdam, The Netherlands; loet@leydesdorff.net

**Abstract**
Strongly anticipatory systems—that is, systems which use models of themselves for their further development—and which additionally may be able to run hyperincursive routines—that is, develop only with reference to their future states—cannot exist in *res extensa*, but can only be envisaged in *res cogitans*. One needs incursive routines of *cogitantes* to instantiate these systems. Unlike historical systems (with recursion), these hyper-incursive routines generate redundancies by opening horizons of other possible states. Thus, intentional systems can enrich our perceptions of the cases that have happened to occur. The perspective of hindsight codified at the above-individual level enables us furthermore to intervene technologically. The theory and computation of anticipatory systems have made these loops between supra-individual hyper-incursion, individual incursion (in instantiation), and historical recursion accessible for modeling and empirical investigation.
**Keywords:** meaning processing; scholarly discourse; communication; sociology.

## 1   Introduction

In a series of lectures in Paris entitled *Cartesian Meditations* (1929), the German philosopher Edmund Husserl formulated as follows:

> We must forgo a more precise investigation of the layer of meaning which provides the human world and culture, as such, with a specific meaning and therewith provides this world with specifically "mental" predicates (Husserl, 1929, at p. 138; my translation).

This layer of meaning—that Husserl would like to study further, but he was aware that the instruments for studying this empirically were failing at the time—consists in his terminology of transcendental, that is, supra-individual, *intentionality*. In a later book entitled *The Crisis of the European Sciences* (1935), Husserl elaborated that this crisis finds its origin in the prevailing belief in the objectivity of science which fails to reflect on the intersubjectivity. According to Husserl, the intersubjectivity *grounds* the objectivity.

"Intersubjectivity" was formulated as a normative criterion for validation in the philosophy of science (e.g., Popper, 1935), but hitherto hardly made a subject of modeling in empirically oriented science studies. How can the social scientist analyze, measure, and model intentionality—the communication of meaning among people—at the supra-individual level? How can the communication of knowledge as a more specific since theoretically rationalized meaning be studied?

Knowledge may lead to a change of the meaning that one can attribute with hindsight to previous events; the facts can be provided with meaning from a new and different perspective. How can one understand the different communication processes of information, meaning, and knowledge, and their possible interactions? I shall argue that the theory and computation of anticipatory systems have provided us with means and methods to understand these processes as different in terms of recursions, incursions, and hyperincursions (Dubois, 1998); these new insights can be elaborated into empirical research about systems which generate redundancies instead of (Shannon) entropy (Leydesdorff, 2010a).

## 2 The communication of meaning and information

Using Shannon (1948, at p. 3) provided a lucid distinction of meaning from information at the beginning of his paper entitled *A Mathematical Theory of Communication*:

> Frequently the messages have *meaning*; that is, they refer to or are correlated according to some system with certain physical or conceptual entities. These semantic aspects of communication are irrelevant to the engineering problem. The significant aspect is that the actual message is one *selected from a set* of possible messages.

Two systems of reference are distinguished in this quotation: the formal one of the electrical engineer or, in other words, Shannon himself as a mathematician, and the possible meanings provided by substantive discourses in which "physical or conceptual entities" can be defined. Shannon considers the latter as irrelevant for the definition of information, but he notes the significance of potentially different *selection mechanisms*.

Shannon-type information is defined as *yet* content-free (Theil, 1972). This condition of "still-to-be-provided-with-meaning" by a system of reference is also manifested in the units of measurement (e.g., bits of information) which are dimensionless. Shannon-type information does not yet contain meaning other than the *mathematical* definition of the expected information value contained as uncertainty in a message as a finite series of differences—in other words, a probability distribution. What the expected information content of the distribution means can only be defined by an observing system using its own selection mechanism. Meaning is defined "in use" (Wittgenstein, 1953).

The meaning provided to the (Shannon-type) information by a system can sometimes reduce uncertainty in this system. Reduction of uncertainty can be measured as negentropy (Brillouin, 1962): this possibility originates from the difference which the difference (or a series of differences, that is, a probability distribution) can make for a receiving system. Thus, "a difference which makes a difference" (Bateson, 1972, p. 489) can only reduce the uncertainty that prevails and be identified as meaningful or observed information, given the specification of a system of reference. The expected information content of the messages, however, necessarily add to the uncertainty (Theil, 1972).[1]

Weaver (1949, at p. 116) noted that Shannon's abstract definition of information as uncertainty sounds "bizarre," but that this level of abstraction might also be needed to develop a theory of meaning. Meaning is generated in use by specific systems that are able to receive and/or process information. This receiving system, however, does not have to be an observer, but can also be a discourse. Information is then provided with meaning which may contain a coordinating function at the supra-individual level. In other words, the meaning is codified *among* human beings, that is, at the intersubjective level.

The systems-theoretical tradition has mainly focused on the observer as the individual unit of analysis (e.g., Edelman, 1989). For example, Von Foerster (1979) ascribed to Maturana as his "Theorem Number One" that "Anything said is said *by* an observer," and added his own corollary that "Anything said is said *to* an observer." He concluded that the two observers share a language. From this perspective, however, language is considered only as a meta-biological domain.

Maturana (1978, at p. 49), for example, noted that for an observer a "second-order consensual domain […] becomes indistinghuishable from a semantic domain." However, understanding in language (as both observers and participants) was set aside by him as the language of a human "super-observer" (pp. 58f.). From this biological perspective, languaging—linguistic behavior—can be observed. The dynamics of human language as different from communication among insects cannot properly be analyzed from these meta-biological perspectives (Epstein and Axtell, 1996).

Human language—and more generally, communication—connects not only observers, but also their observational reports, that is, the *translation of their observations into communication* provides the messaging with intersubjective meaning, and this codification allows for the communication of meaning at the supra-individual level (Mead, 1934; Pask, 1975). Distinguishing the observational reports in language from the observers making only "utterances"[2] moves us from the realm of mathematical biology and the behavioral sciences into the realm of communication as interhuman, that is, meaningful understanding and potentially knowledge-based coordination.

## 3   Language and symbolic mediation

Meaning is generated in a system when different pieces of information are related as messages to one another, for example, as words in sentences (Hesse, 1988; Law & Lodge, 1984). The information is then positioned in a network with an emerging (and continuously reconstructed) structure. This positioning can be done by an individual

---

[1] The second law of thermodynamics holds equally for probabilistic entropy, since $S = k_B H$ and $k_B$ is a constant (the Boltzmann constant). Because of the constant, the development of $S$ over time is a function of the development of $H$, and *vice versa*.

[2] Luhmann ([1984], 1995) defined communication in terms of three elements: (1) utterance, (2) information, and (3) understanding. Such a definition, in my opinion, does not distinguish sufficiently between communication among animals (e.g., insects) and human beings (Leydesdorff, 2006).

who—as a system of reference—can provide subjective meaning to the events, but meaning can also be provided at the supra-individual level, for example, in a discourse. In the latter case, meaning is discursive, and its dynamics can therefore be expected to be different from those of psychological meaning(s).

Whereas a psychological identity can be expected to strive to integrate a plurality of meanings that could be provided to single events (for example, in order to avoid "cognitive dissonance"), the social system—as a potentially decentered *dividuum* (Nietzsche, 1878: I, §57)—can tolerate the entertainment of different meanings, and has the additional option of being differentiated into subsystems which codify these meanings according to their own standards. This plurality in rationalities can be considered functional to the processing of complexity in a pluriform society (Bourdieu, 2004).

For example, politicians and economists can discuss "shortages of energy" although among physicists "energy" is considered as a *conserved* quantity. Codifications facilitate and speed up the communication by making the communication system-specific. At the market, for example, one can simpy pay the price of something without having to negotiate. The price codifies the value of the commodity. Prices make it possible to abstract from the underlying values in another semantic domain (for example, that of banking).

This possibility of functional differentiation in the codes of communication and the potentially symbolic generalization of meaning was first elaborated in the tradition of social-systems theory by the sociologists Talcott Parsons and Niklas Luhmann. Building on Durkheim's argument that norms function as integrative at the supra-individual level, Parsons (1961) theorized the possibility of functional differentiation among the roles of agents in different subsystems of society. Following Merton (1957), Luhmann ([1984], 1995) historicized social functions and proposed that they develop as specific rationalities in the different and historically variable processings of meaning in social subsystems. Luhmann added the evolutionary perspective that new forms of codification can be invented; for example, at first coins were used, then banknotes, and much later credit cards. Each communication subsystem develops further by overwriting and repositioning the previous versions of its coding.

The prime example of this cultural evolution of communication has been provided by Kuhn's (1962) notion of paradigm shifts. "Phlogiston," for example, was backgrounded in scholarly discourse once "oxygen" was constructed as a new concept (Priestly, 1774-1777). The new paradigm (in chemistry) opened domains for puzzle-solving and further communication with a code that is different from the previous one. However, both "oxygen" and "phlogiston" use a code of communication very different from exchange processes on the market, which obey economic mechanisms of exchange. Analogously, the truth of a scientific statement is different from the religious truth of a dogma. Other dimensions of interhuman communication (e.g., affection, power) always also play a role, although the institutional setting may facilitate the functionality of specific codes among these symbolically generalized media more than others. For example, it is transgressive to favour one scientific theory over another for political reasons, or to bribe a judge.

Interhuman communication can thus be considered as a fabric woven in many directions: each communication can be provided with meaning in terms of power, economic utility, affection, scientific truth, etc. These latent dimensions of communication resound and operate selectively in all interhuman communication. However, in specific communications some selections can be expected to operate more strongly than others because of the functionality of coding. Symbolically generalized codes enable us to be specific in our communications and thus to process more complexity.

The selecting codes of communication are not a given, but enacted and reconstructed in use as the culturally and therefore supra-individually constructed dimensions of communication. Note that the codes are attributes to the communication and not the (human) communicators. Communication, however, is not like agency an identifiable unit of analysis, but an operation which remains in flux.

Herbert Simon (1973) hypothesized that any evolving system can be expected to operate with an alphabet. Thus, one might hypothesize 20+ symbolically generalized media of communication possible in interhuman communications. These codes of communication should not be reified: they are historically constructed and enacted bottom-up in interhuman communications, but as they are reconstructed recursively over time, they can be expected to function as incursive control mechanisms at the level of society that enable us to enrich our communication by allowing for greater precision.

The codes operate as selection mechanisms by enabling us to focus the communication. Selection mechanisms can reinforce one another in processes of mutual shaping (McLuhan, 1964). Thus, selection mechanisms can be expected to shape historical trajectories that are relatively stable (for example, in institutions). A next-order selection may drive a local stabilization into global meta-stabilization, or into regimes which function with dynamics that differ from—since they counteract as feedback mechanisms on—the dynamics of historical developments. Such further differentiation among selection mechanisms (stabilization and globalization) can uncouple the communication reflexively from the historical process in which it emerges.

For example, communication with money first speeds up the communication on the market to the extent that local forms of capitalism can be shaped. Bank notes, stock exchanges, and credit cards provide means for worldwide transactions with correspondingly increasing speeds and precision. Marx already identified this emerging mechanism in capitalism as "alienating." Luhmann ([1984], 1995) proposed to study the dynamics of communication (cf. Marx's "exchange value") as analytically different from the dynamics of human or group behavior ("use value"). The systems of reference are altered by the change of perspective caused by the newly emerging code of communication. This potential globalization is an attribute of the communication and not of the communicators.

In the terminology of *autopoiesis* theory (Maturana & Varela, 1980), the two dynamics of processing meaning—at the level of agency and at the social level—remain "structurally coupled" and "interpenetrate" each other reflexively (Luhmann, 2002). However, one can expect the cybernetics of communication to be different from the dynamics of human (group) behavior. Communications, for example, can travel worldwide without the communicators as carriers having to move.

## 4 Horizons of meaning

Whereas the biological *autopoiesis* processes the history of the communications in terms of their structural sediments—for example, in terms of differentiations among organs or species—an orientation toward horizons of possible communication is provided by the additional communication of meaning (denotations and connotations) in language. The linguistically or symbolically mediated communication channels are changed by the historical communication in terms of the communications *possible* thereafter. In other words, redundancies—sets of possibilities—are continuously generated. Unlike the biological *autopoiesis* of the living (Maturana & Varela, 1980), meaning can be communicated reflexively and with reference to and in anticipation of "horizons of meaning." I use the plural of this Husserlian concept in order to emphasize that one can expect the horizons of meaning to be structured by symbolically generalized codes of communication which are functionally different.

Interhuman communication is based on interactions among both recursive communicators and incursive communications. The meanings interact in a non-linear dynamics which is not hardwired and therefore no longer necessarily subject to the second law of thermodynamics. The redundancies generated by the processing of meaning—with reference to other possible meanings—can structure and reconstruct the information processing from the perspective of hindsight. The possibility of such non-linear dynamics is enabled by language as an evolutionary achievement: meaning can proliferate discursively at a speed much faster than its instantiations in language (e.g., in fantasies and wishes) because of the possible feed-forward loops between individual experiences and expectations, and communication in language.

Luhmann (1986) criticized Schutz's (1932; 1953) interpretation of Husserl's phenomenology because of the emphasis on observable instantiations of the cybernetics of behavior in the *life-world* (cf. Habermas, 1981). Husserl himself had formulated his philosophy as a "transcendental phenomenology" with an emphasis on expectations (as different from and richer than observations). From the perspective of hindsight, this focus on intentions, meanings, and expectations can be considered as "mathematical" in the sense that it enables us to intuit other possible dimensions which resound in the empirical events that have happened to occur historically (Derrida, 1974; Husserl, [1935/6] 1962). We have no access to the possible other than by placing what exists reflexively between brackets. Husserl used the Greek word epochè (ἐποχη) for this "suspension" of all judgments about the existence of an external world.

The analytical specification of expectations before proceeding to actual observations enables us to specify whether observed differences (variation) can be considered as significant. From this "phenomenological" perspective, the phenomena are enriched with other possible options. This is formalized, for example, in the chi-square test of statistical significance using a theory of measurement. Husserl (1935/6) noted that the positivistic focus on observables had eroded this epistemological basis of the modern sciences, and that one should instead return to the reflexive position of Descartes, but reconstruct it in order to ground the (social) sciences in the intentional reflexivity of human communications.

Not incidentally, therefore, Husserl (1929) called the book in which he explained his intersubjective "horizons of meaning" *Cartesian Meditations*. Husserl's reference is to Descartes' distinction between *res extensa* and *res cogitans*. The Cartesian *Cogito* knows him/herself as uncertain and different from the external world. In this act of doubt, the contingent *Cogito* finds the transcendent environment as the Other or a personal God. However, Husserl doubted this next step: the *cogitatum* of the *Cogito* is not necessarily God, but can also be considered as an intersubjective domain to which we all have personal access: the horizons of meaning that we share (to different extents). This domain is not in the *res extensa*, but remains *res cogitans*. In other words, the meaning that we provide to the events does not "exist" physically, but incurs on us as one among a set of culturally possible meanings.

In the social sciences—as in the other theoretical sciences—one can use models to specify the expectations. The specification of expectations makes future states available in the present as potentially meaningful. The model—as different from the individual intuition—enables us to communicate about these future states with greater precision by invoking the symbolic codes of scholarly discourse. A model thus is part of discursive knowledge; it can be improved by argumentative contributions. However, this providing of new meaning is highly codified; only those who understand the model can contribute meaningfully. The model enables these participants to entertain a communication among specialists in which further knowledge can be developed and exchanged. In other words, the model is part of a communicative reality.

This communicative reality that the communicators shape over time and reflexively reconstruct cannot be considered as *res extensa*, but belongs to the *res cogitans*; it is not stable like matter, but remains in flux like language. Language enables us to communicate in terms of uncertainties (e.g., possibly relevant questions) and expectations. Husserl (1929) recognized this realm as *cogitatum*, that is, the substance about which the *Cogito* remains uncertain. Our mental predicates provided to the world in intersubjective exchanges with intentional human beings, shape our culture and therewith ground what Husserl also called a "concrete ontology" or, in other words, "a universal philosophy of science" (1929, at p. 159).

This philosophy of science enables us to understand scientific models and concepts as specifically coded meanings that we attribute to an external reality at the intersubjective level. Note that from this perspective, the external world is not a social construct as in post-modernism; it is a cognitively hypothesized and highly codified construct that can be accessed reflexively and perhaps partly reconstructed by agency, but only in terms of further communications. In the *res extensa*, resources can be mobilized (for example, at the institutional level), but such policies can succeed only insofar as they enable us to access, deconstruct, and reconstruct the self-organization of cultural constructs in the *res cogitans*.

The models and not the modeled substances shape the sciences as cultural artifacts. According to Husserl's (1935) *The Crisis of the European Sciences*, however, an empiristic self-understanding prevails in the modern sciences. In order to move the (social) sciences forward, one has to stay with the transcendental that one can retrieve in one's self (incursively) and thus recognize the sciences as part and parcel of a (hyperincursive) realm of cultural expectations that can be communicated. The "naturally" observed or perceived at the individual level is shaped by and rewritten in a realm of intersubjective expectations and their possible communication. Note that this *res cogitans* is also *res*, that is, part of a reality, and thus a possible subject of empirical investigation. The predictions on the basis of the models, for example, can be expected to feed back (e.g., technologically) on our material life.

## 5    The structuration of expectations

The scientific model as an exchange mechanism of cognitive expectations can provide us with a heuristics to understand the communication of other, for example, normative expectations. Normative exchanges can be expected to shape, for example, political discourse. In political discourse, events are provided with meanings that differ from those given by scientific discourse. In other words, social order is not a given, but a set of variously codified expectations that interact and self-organize in the *res cogitans*. This order of expectations can be sustained by institutions (which function as instantiations; cf. Giddens, 1984). Political discourse, for example, can be focused in a parliamentary debate, whereas scholarly discourse can be retrieved in scientific journals. The complex and internally differentiated order of expectations remains latent. The manifest instantiations can also be considered as their co-variations in mixtures at specific moments of time.

The codes can be considered as the latent dimensions that structure the discourses in analytically different directions. Two levels can be distinguished in this structuration. First, the codes operate over time as the internal axes of meaning-providing structures; for example, in natural language. Second, this operation is recursive: meaning is provided to the information contained in the events, and meaning can further be codified—that is, provided with symbolic meaning—in the communication. However, this requires a functional language such as a scientific discourse Under the condition of functional differentiation the axes can be expected to span horizons of meanings in increasingly orthogonal directions. Some codes can in a next selection be generalized symbolically or, from the perspective of their stabilization along trajectories, be globalized as horizons of meaning that feed back on the local meaning processing in interhuman communication.

Luhmann (1986) provided an elaboration of Husserl's concept of "intersubjectivity" in sociological terms. Three levels were distinguished in the communication of meaning: (*i*) local interactions, (*ii*) organization of meaning in historical instances, and (*iii*) the self-organization of the codes of communication. In modern, pluriform societies self-organization in different directions can be expected to prevail—communications are no longer coordinated at the center—while in pre-modern societies communication was organized in terms of institutions. Organization integrates, while self-organization tends to differentiate the functions of the communication. Organization operates at specific moments of time and self-organization of meaning operates over time as codification in fluxes of communication.

The two cybernetic mechanisms of organization (integration at interfaces in the present) and self-organization (differentiation of codes) can be considered as the incursive woof and hyperincursive warp of the evolution of the *cogitatum* in a multidimensional space. This evolutionary development is driven bottom-up by variation in the interactions, while the codes operate in terms of selection mechanisms. For example, economic exchanges are organized in terms of local markets, but can self-organize a global market equilibrium if left sufficiently free to do so. Scientific communication is organized within communities and institutions, but these communities compete in hypothesizing and following the dynamics at the level of scientific fields (Bourdieu, 1976, 2004). Love and affection can be organized in terms of marriages, but also otherwise (Luhmann, 1982). Organization of meaning is historically contingent, and can thus be studied empirically.

## 6  The processing of meaning and the computation of anticipation

I noted above that meanings incur on events. In other words, events are understood from the perspective of hindsight, but with reference to possible future events. The time axis is thus a crucial dimension. It is often pictured as an arrow flying from the past via the present to the future, but the retrospective perspective of hindsight also operates in the processing of meaning, and thus time can be considered as yet another dimension or a degree of freedom. A model can provide us with a prediction about future states because it remains in a *res cogitans* in which future moments of time can be simulated in the present.

The mathematical biologists Rosen (1985) defined as "anticipatory" a system that is able to entertain a model of itself. The model can provide the system with one or more representations of future states in the present. These representations can be used for the active reconstruction of the system. Dubois (2003, at pp. 112f.; 2006, at pp. 59f.) further distinguished between weakly and strongly anticipatory systems: a system that is weakly anticipatory is able to use its *predicted* states at future times for adaptation or intervention, while a system which is strongly anticipatory can use its anticipated states at time $t + 1$, $t + 2$, etc., for its present reconstruction.

Furthermore, Dubois (1998) distinguished between incursive and hyper-incursive systems. Incursive systems use both their historical states and present or future states for their reconstruction, whereas hyper-incursive systems operate exclusively on the basis of expectations. The *cogitans* is embodied and therefore uses also historical states for the computation of a next one (in the present), whereas the social system or Husserl's *cogitatum* can further be developed on the basis of possible future states, that is, expectations and their organization in systems of expectations. Such hyper-incursive cybernetics would operate against the axis of time and thus reduce uncertainty. Note that this is not yet to be considered as a system, but a mechanism which requires anchoring in historical time by other (incursive) mechanisms.

In other words, a *cogitatum* cannot be instantiated without a *cogitans* to instantiate it; the hyperincursive *cogitatum* (horizon of expectations) and the incursive *cogitans* are structurally coupled and intertwined by reflexive interpenetration. Thus, there is always historical production of uncertainty in the "life-world" involved, but this forward arrow is counteracted by a feedback arrow from the self-organization of the codes. The evolutionary mechanism operates against the arrow of time because redundancies—other possibilities—are generated instead of Shannon-type information. More technically, one can formulate that the maximum information content can be expected to expand more rapidly than the expected information content, and thus redundancy—reduction of uncertainty—is the net result (e.g., Brooks & Wiley, 1986).

A hyper-incursive system cannot "exist" and be observed in the *res extensa*; it remains a possibly functioning routine that enriches our understanding of the observable reality. The definition of it is analytical and should not be reified in an external world. In other words, these are relevant subdynamics for the specification of the dynamics of communication of meaning and knowledge. However, the domains to be studied from this perspective are very different from biological or even psychological ones (Giddens, 1979). For example, it could be shown—using the logistic equation (which can be used for modeling processes of growth and decline in biology) and its equivalent formulation in the hyper-incursive domain (Dubois, 1998)—that the biological and sociological domains are separated at the value of four of the so-called bifurcation parameter, and that this separatrix can only be crossed by invoking an incursive routine, that is, a psychological *cogitans* or human agency (Leydesdorff & Franse, 2009). Human beings having both a body and a mind are able to participate in both domains.

The three cybernetic mechanisms specified by Luhmann can be operationalized as *different* mechanisms of incursion and/or hyper-incursion. Let me not repeat the derivation of the various equations (Leydesdorff, 2008, 2010b). When the various equations are solved, the conclusions are the following:

First, Dubois' (1998) formulation and derivation of the hyperincursive formulation of the logistic equation [ $x_t = ax_{t+1}(1 - x_{t+1})$ ] can be provided with an interpretation as reflexivity about each other's expectations in interpersonal relations (Parsons, 1968): *Ego* ($x_t$) in the present can further develops his/her expectations in terms of its own expected next states ($x_{t+1}$) and with reference to the next states of his/her environment $(1 - x_{t+1})$. However, a decision is needed for the instantiation.

In a second step (Leydesdorff, 2008), one can abstract from personhood and self-reference in the case of studying social systems, and focus on the selections among communications as selection mechanisms of next possible states. One can then distinguish (Luhmann's) three mechanisms and three corresponding equations:

*Interaction*—hyper-incursively modeled as the interaction of two hyperincursive selection mechanisms [ $x_t = b\,(1 - x_{t+1})(1 - x_{t+1})$ ] leads to turn taking in the communication of meaning and thereby variation from the perspective of the social communication system.

By extending this model parsimoniously only with a single (third) selection mechanism, one obtains two options which model hyper-incursively *self-organization*

**182**

and incursively *organization* of meaning, respectively. When three anticipatory mechansims can operate selectively upon one another, this could be modeled as follows:

$$x_t = c\ (1 - x_{t+1})(1 - x_{t+1})(1 - x_{t+1}) \tag{1}$$

Organization instantiates the interfaces historically, and one can analogously model this as follows:

$$x_t = d\ (1 - x_{t+1})(1 - x_{t+1})(1 - x_t) \tag{2}$$

The difference between the organization of meaning (eq. 2) and its self-organization (eq. 1) is provided in the third term: is the processing bent back to the present, or does this term in the model remain a reference to a future state? Organization can reduce uncertainty by instantiation in the present.

Eq. 1 (modeling *self-organization*) has two imaginary roots and one real root (see Leydesdorff (2008) for the derivation). The real root can be considered as a constant operating in the coding; and used by the strongly anticipatory system for maintaining its hyper-incursive "identity." Eq. 2 models a system (*organization* of meaning) that can be expected to perish historically after a finite number of instantiations. In other words, organizations of meaning emerge and disappear historically. Long-term codification is provided to the communication meta-historically by the evolutionary mechanism of self-organization in the communication of meaning. The third mechanism—interaction—provides variation. Organization and self-organization are coupled to each other as eqs. 1 and 2, but the results of these two mechanisms operating provide different (integrating and differentiating) solutions.

## 7   Conclusion

I have argued that symbolically generalized media of communication codify our expectations and thereby empower our performances reflexively in terms of handling complexity in terms of expectations. The reflexive understanding of horizons of meaning made possible by communication provides us with access to a social reality in which knowledge-based anticipations play an increasing role. The (self-)organization of meaning at the above-individual level no longer "structurates" only our *actions* (Giddens, 1979, 1984), but more importantly our *expectations*.

The theory and computation of anticipatory systems has provided us with access to a domain which has been anticipated in Husserl's philosophy. Vice versa, this philosophical reflection makes us aware that one can expect no natural systems to be hyperincursive other than *cogitata*. The horizons of expectations are accessible only to *cogitantes* who in turn are needed for the incursive instantiations of expectations. Perhaps, the algorithms may in the longer run be useful for the construction of engineered systems that support man-machine interactions by extending our domains of expectations.

First, pieces of information can be related and povided with meaning. In biological systems, this mechanism is fixed (as connotations) such as in the case of stimulus-response reactions. The communication of meaning in terms of both denotations and connotations, however, can self-organize a next-order layer because stability in terms of a "self" is no longer required for the cultural reproduction of communication. Horizons of meaning thus open up at the top of the cultural evolution which enable us to model, communicate the models, and technologically to intervene on the basis of specifications of expectations.